\documentclass[12pt]{article}

\usepackage{graphics}

\title{Velocity Autocorrelation and Harmonic Motion in Supercooled 
       Nondiffusing Monatomic Liquids}
\author{Eric D.\ Chisolm, Brad E.\ Clements, and Duane C.\ Wallace \\ 
        Theoretical Division \\ Los Alamos National Laboratory \\ Los Alamos, 
        NM~~87545}

\begin{document}

\maketitle

\begin{abstract}
Studies of the many-body potential surface of liquid sodium have shown that it 
consists of a great many intersecting nearly harmonic valleys, a large fraction
of which have the same frequency spectra.  This suggests that a sufficiently 
supercooled state of this system, remaining in a single valley, would execute 
nearly harmonic motion.  To test this hypothesis, we have compared 
$\hat{Z}(t)$, the normalized velocity autocorrelation function, calculated 
from MD simulations to that predicted under the assumption of purely harmonic 
motion.  We find nearly perfect agreement between the two, suggesting that the 
harmonic approximation captures all essential features of the motion.
\end{abstract}

\section{Introduction}
\label{intro}

Recent work by Wallace and Clements \cite{enkin, radang} has uncovered several 
important properties of the many-body potential underlying the motion of  
liquid sodium systems.  Specifically, it has been shown that (a) the potential 
surface consists of a large number of intersecting nearly harmonic valleys, (b)
these valleys can be classified as symmetric (crystalline, microcrystalline, or
retaining some nearest-neighbor remnants of crystal symmetry) or random, with 
the random valleys vastly outnumbering the symmetric ones, (c) the frequency 
spectra of different random valleys are nearly identical (while those of the 
symmetric valleys vary widely), and (d) below 35 K the system remains in a 
single valley throughout the longest molecular dynamics (MD) runs that were 
performed.  Results (a) through (c) verify predictions made by Wallace 
in his theory of liquid dynamics \cite{liqdyn}, which has been successfully 
applied to account for the high-temperature specific heats of monatomic liquids
\cite{specif} and a study of the velocity autocorrelation function 
\cite{oldvacf}.  These four results together suggest that below 35 K the motion
of the atoms in liquid sodium is purely harmonic to a high degree of 
approximation, again as predicted by Wallace in \cite{liqdyn}, and we would 
like to test this hypothesis further.  One check is to compare the mean square 
displacement from MD with the prediction from purely harmonic motion, which is 
done in Fig.\ 12 of \cite{enkin}, where the two are found to agree closely.  
However, it would be more convincing if the theory could be shown to reproduce 
an entire scalar function calculated from MD (instead of just a single number),
such as the normalized velocity autocorrelation function $\hat{Z}(t)$.  That is
the aim of this paper.  We will show that purely harmonic motion of the atoms 
in a potential valley produces a $\hat{Z}(t)$ which matches that of MD 
calculations to within the calculations' accuracy; thus we will conclude that 
the motion of atoms in a nondiffusing supercooled liquid state is very nearly 
entirely harmonic.  For completeness, in Sec.\ \ref{theory} we briefly review
the calculation of $\hat{Z}(t)$ assuming harmonic motion, and in Sec.\ 
\ref{MD} we compare this result with MD.  Finally, in Sec.\ \ref{concl} we make
contact with work by others in this field, as well as comparing these results 
to Wallace's earlier effort \cite{oldvacf} mentioned above.

\section{Harmonic Theory}
\label{theory}

If an $N$-body system is moving in a potential valley, the potential can be 
expanded about the valley minimum with the resulting Hamiltonian
\begin{equation}
H = {\sum_{Ki}}' \, \frac{p_{Ki}^{2}}{2M} + {\sum_{Ki,Lj}}' \, \Phi_{Ki,Lj} 
    u_{Ki} u_{Lj} + \Phi_{A}
\label{Hamiltonian}
\end{equation}
where $u_{Ki}$ is the $i$th component of the $K$th particle's displacement 
from equilibrium, $p_{Ki}$ is the corresponding momentum, and the anharmonic 
term $\Phi_{A}$ contains all of the higher order parts of the expansion.  The 
primed sum indicates that the sum is performed under the constraint that the 
center of mass of the system is stationary.  (As a result, the system has only 
$3N-3$ independent degrees of freedom.)  The matrix $\Phi_{Ki,Lj}$ is called 
the dynamical matrix of the system.  If the valley is approximately harmonic, 
we can neglect $\Phi_{A}$.  If coordinates $q_{\lambda}$ are defined by the
relation 
\begin{equation}
u_{Ki} = \sum_{\lambda} w_{Ki, \lambda} q_{\lambda}
\label{newcoords}
\end{equation}
where the $w_{Ki, \lambda}$ form a $3N \times 3N$ orthogonal matrix, 
satisfying 
\begin{equation}
\sum_{Ki} w_{Ki, \lambda} w_{Ki, \lambda'} = \delta_{\lambda\lambda'},
\label{orth}
\end{equation}
then the Hamiltonian in these new coordinates is
\begin{equation}
H = \sum_{\lambda} \frac{p_{\lambda}^{2}}{2M} + {\sum_{Ki,Lj}}' 
    \sum_{\lambda\lambda'} w_{Ki, \lambda} \Phi_{Ki,Lj} w_{Lj, \lambda'} 
    q_{\lambda} q_{\lambda'}
\end{equation}
where the $p_{\lambda}$ are the momenta conjugate to the $q_{\lambda}$.  Now 
one can always choose the $w_{Ki, \lambda}$ to diagonalize $\Phi_{Ki,Lj}$, so 
that
\begin{equation}
\sum_{Ki,Lj} w_{Ki, \lambda} \Phi_{Ki,Lj} w_{Lj, \lambda'} = 
               M \omega_{\lambda}^{2} \delta_{\lambda\lambda'}.  
\label{diagphi}
\end{equation}
(This equation defines the frequencies $\omega_{\lambda}$ in terms of the 
eigenvalues of $\Phi_{Ki,Lj}$.)  With this choice, the Hamiltonian becomes
\begin{equation}
H = \sum_{\lambda} \left( \frac{p_{\lambda}^{2}}{2M} + 
                   \frac{1}{2}M\omega_{\lambda}^{2}q_{\lambda}^{2} \right).
\end{equation}
Three of the $\omega_{\lambda}$ are zero; these modes correspond to uniform 
motion of the center of mass.  Since we have restricted the center of mass 
position and velocity to zero, these modes are not excited.  The classical 
equations of motion for the remaining modes are solved by
\begin{equation}
q_{\lambda}(t) = a_{\lambda} \sin(\omega_{\lambda}t + \alpha_{\lambda}),
\end{equation}
or, returning to the original coordinates,
\begin{equation}
u_{Ki}(t) = \sum_{\lambda} w_{Ki, \lambda} a_{\lambda} \sin(\omega_{\lambda}t +
            \alpha_{\lambda}),
\label{uoft}
\end{equation}
with the understanding that the sum on $\lambda$ ranges from $1$ to $3N-3$.
The velocities of the particles are
\begin{equation}
v_{Ki}(t) = \sum_{\lambda} w_{Ki, \lambda} \, \omega_{\lambda} a_{\lambda} 
            \cos(\omega_{\lambda}t + \alpha_{\lambda}).
\label{voft}
\end{equation}
We compute the $\langle \mbox{\boldmath $v$}(t) \cdot \mbox{\boldmath $v$}(0) 
\rangle$ in $Z(t)$ by calculating $\mbox{\boldmath $v$}_{K}(t) \cdot 
\mbox{\boldmath $v$}_{K}(0)$, summing over $K$ and dividing by $N-1$ (remember 
that only $3N-3$ coordinates are independent), and averaging over the 
amplitudes $a_{\lambda}$ and phases $\alpha_{\lambda}$.  Thus
\begin{eqnarray}
Z(t) & = & \frac{1}{3} \langle \mbox{\boldmath $v$}(t) \cdot \mbox{\boldmath 
           $v$}(0) \rangle \nonumber \\
&   & \nonumber \\
& = &  \frac{1}{3N-3} \sum_{Ki} \sum_{\lambda\lambda'} w_{Ki, \lambda} w_{Ki, 
       \lambda'} \omega_{\lambda} \omega_{\lambda'} \langle a_{\lambda} 
       a_{\lambda'} \rangle \langle \cos(\omega_{\lambda}t + \alpha_{\lambda}) 
       \cos(\alpha_{\lambda'}) \rangle \nonumber \\
 &   & \nonumber \\
 & = & \frac{1}{3N-3} \sum_{\lambda\lambda'} \delta_{\lambda\lambda'}
       \omega_{\lambda} \omega_{\lambda'} \langle a_{\lambda} a_{\lambda'} 
       \rangle \langle \cos(\omega_{\lambda}t + \alpha_{\lambda}) 
       \cos(\alpha_{\lambda'}) \rangle \nonumber \\ 
 &   & \nonumber \\
 & = & \frac{1}{3N-3} \sum_{\lambda} \omega_{\lambda}^{2} \langle 
       a_{\lambda}^{2} \rangle \langle  \cos(\omega_{\lambda}t + 
       \alpha_{\lambda}) \cos(\alpha_{\lambda}) \rangle \nonumber \\
 &   & \nonumber \\
 & = & \frac{1}{6N-6} \sum_{\lambda} \omega_{\lambda}^{2} \langle 
       a_{\lambda}^{2} \rangle \cos(\omega_{\lambda}t).
\label{vdotv}
\end{eqnarray}
By the equipartition theorem, 
\begin{equation}
\left\langle \frac{1}{2} M \omega_{\lambda}^{2} q_{\lambda}^{2} \right\rangle =
         \frac{1}{2}kT
\end{equation}
for any nonzero $\omega_{\lambda}$, from which it follows that
\begin{equation}
\langle a_{\lambda}^{2} \rangle = \frac{2kT}{M\omega_{\lambda}^{2}},
\label{equipart}
\end{equation}
so
\begin{equation}
Z(t) = \frac{1}{3N-3} \frac{kT}{M} \sum_{\lambda} \cos(\omega_{\lambda}t).
\label{Z}
\end{equation}
Notice that $Z(0) = kT/M$, so $\hat{Z}(t)$ defined by 
\begin{equation} Z(t) = Z(0) \hat{Z}(t) \end{equation}
is given in this theory by
\begin{equation}
\hat{Z}(t) = \frac{1}{3N-3} \sum_{\lambda} \cos(\omega_{\lambda}t).
\label{Zhat}
\end{equation}
This is the result we wish to compare with MD.

To do so, we need the frequencies $\omega_{\lambda}$, which are related to the
eigenvalues of the dynamical matrix $\Phi_{Ki,Lj}$ as indicated in Eq.\ 
(\ref{diagphi}).  These were evaluated for five separate random valleys in 
\cite{enkin} by quenching all the way down to a valley minimum and 
diagonalizing $\Phi_{Ki,Lj}$ there; as pointed out in Sec.\ \ref{intro}, these 
eigenvalues were found to be independent of the specific random valley chosen. 
All five sets of eigenvalues are shown in Fig.\ 7 of \cite{enkin}, and we 
picked one set at random to use in performing the sum in Eq.\ (\ref{Zhat}); the
other sets produce identical graphs of $\hat{Z}(t)$.  

\begin{figure}
\includegraphics{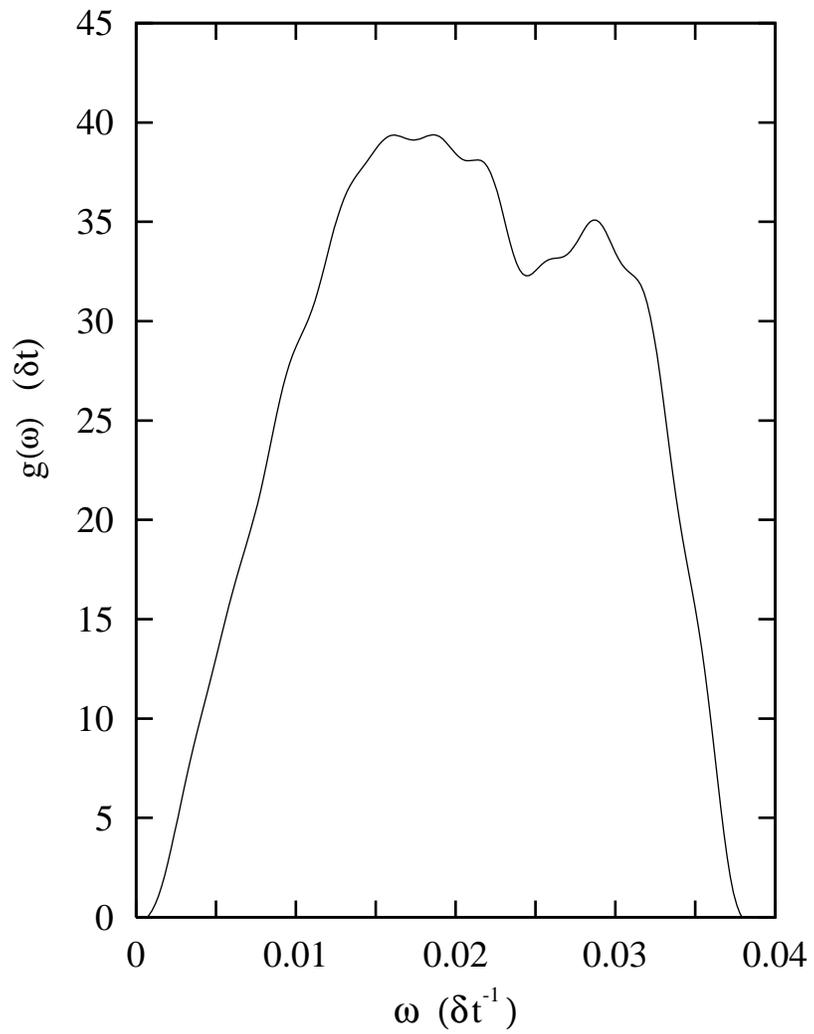}
\caption{$g(\omega)$ constructed from one of the five sets of frequencies in 
         Fig.\ 7 of \cite{enkin}.  This same set of frequencies is used to 
         calculate $\hat{Z}(t)$ from Eq.\ (\ref{Zhat}).}
\label{nvsfreq}
\end{figure}

We can also use the set of eigenvalues to reconstruct the density of 
frequencies $g(\omega)$; the results are shown in Fig.\ \ref{nvsfreq}.  Note 
that we do not actually integrate over this $g(\omega)$ to
evaluate $\hat{Z}(t)$ below; we directly sum over the given set of frequencies 
as indicated in Eq.\ (\ref{Zhat}).  The Figure is provided only to convey a 
sense of the shape of the frequency distribution.  Also note that this 
$g(\omega)$ is determined from fully mechanical considerations; as a result, it
is not temperature-dependent as are the frequency spectra used in Instantaneous
Normal Mode (INM) studies \cite{RMM, LVS, Stratt, VB, MKS}.  We will expand on 
this point in the Conclusion.

\section{Comparison with MD}
\label{MD}

The MD setup used to calculate $\hat{Z}(t)$ to compare with Eq.\ (\ref{Zhat}) 
is essentially that described in \cite{enkin}:  $N$ particles interact through 
a potential that is known to reproduce accurately a wide variety of 
experimental properties of metallic sodium (see discussion in \cite{enkin} for 
details).  The two significant changes are that we used $N=500$ for all runs 
and that the MD timestep was reduced to $\delta t = 0.2 t^{*}$, where $t^{*} = 
7.00 \times 10^{-15}$ s is the natural timescale defined in \cite{enkin}.  (The
system's mean vibrational period $\tau = 2\pi/\omega_{\rm rms}$, where the rms 
frequency $\omega_{\rm rms}$ is calculated in \cite{enkin}, is approximately 
$300 \,\delta t$.)  We cooled the sodium sample to 22.3 K and 6.69 K, and then 
we ran each at equilibrium to collect velocities $\mbox{\boldmath $v$}_{K}(t)$ 
to be used to calculate $Z(t)$ by the formula
\begin{equation}
Z(t) = \frac{1}{3N} \sum_{K} \frac{1}{n+1} \sum_{t'=0}^{n} 
       \mbox{\boldmath $v$}_{K}(t+t') \cdot \mbox{\boldmath $v$}_{K}(t').
\label{MDZ}
\end{equation}
We then divided by $Z(0)$ to obtain $\hat{Z}(t)$.  The number $n$ was chosen 
as large as possible without running beyond the data calculated in the MD run. 
We know that at these temperatures the sodium is nondiffusing for two reasons: 
Both temperatures are below the 35 K threshold \cite{enkin}, and $\hat{Z}(t)$ 
from either MD run (shown below in Fig.\ \ref{Zvst}) integrates to zero, 
yielding zero diffusion coefficient.

The formula above may fail to produce reliable values of $\hat{Z}(t)$ for three
reasons.  First, the number of data points in the time average may be too 
small; if the MD simulation is run out to time $t_{\rm max}$, then for a given 
value of $t$ in Eq.\ (\ref{MDZ}), the maximum possible value of the upper limit
$n$ is $t_{\rm max}-t$.  Thus we require $t_{\rm max} >> t$; we have chosen 
$t_{\rm max} = 50,000$ timesteps and we have calculated $\hat{Z}(t)$ only to 
$t = 1000$.  To ensure that this value of $t_{\rm max}$ is large enough, we 
also performed MD runs out to 200,000 timesteps and calculated $\hat{Z}(t)$ 
from them; the differences from the 50,000 timestep result were of order 
$10^{-3}$.  Hence we are confident that 50,000 timesteps is enough if we 
calculate $\hat{Z}(t)$ to only 1000 timesteps.  Second, it is possible that 
reducing the timestep (thus increasing the accuracy of the simulation) might 
improve the accuracy of $\hat{Z}(t)$.  To test this, we performed another MD 
run with $\delta t$ reduced to $0.05 t^{*}$, keeping the ``real'' time of the 
run the same; this also produced differences in $\hat{Z}(t)$ of order 
$10^{-3}$.  Thus we are sure that our timestep is small enough.  Finally, there
is the possibility of finite size effects.  Since the MD system has periodic 
boundary conditions, an acoustic wave sent out from the system at $t=0$ could 
propagate across the simulation region and return to its point of origin in a 
finite time, producing spurious correlations that would show up in $\hat{Z}(t)$
but would not be present in a large-$N$ system.  To see if this effect is 
relevant, we estimated the time it would take for an acoustic wave to cross the
region, using the numbers from \cite{enkin}.  The speed of sound in sodium at 
its melting temperature is $2.5 \times 10^{5}$ cm/s, and the volume of the 
region occupied by one atom is $278 \, a_{0}^{3}$, so from the fact that there 
are 500 atoms one finds that the time required for an acoustic wave to cross 
the region is $783 \, \delta t$, or about 800 timesteps.  (The speed of sound 
in sodium at our lower temperatures varies from that at the melting point by 
roughly 5\%, so this result is valid to the same accuracy.)  In the Figure 
below, the MD result for $\hat{Z}(t)$ begins to show small oscillatory revivals
at about this time; we conclude that this is a finite size effect, but it does 
not affect the data before that time. 

In Fig.\ \ref{Zvst}, Eq.\ (\ref{Zhat}) is plotted on top of the MD data for
$\hat{Z}(t)$ of sodium at the two temperatures.  Although both temperatures 
compare exceptionally well to the harmonic theory, the match is visibly poorer 
for the lower temperature.  However, repeated MD runs at the lower temperature 
revealed that overall variations in $\hat{Z}(t)$ amount to $10^{-2}$ on
average, which is the same order as the differences between theory and MD in 
this Figure.  By 500 timesteps the theory is slightly out of phase with the MD 
data, and this small difference persists out to more timesteps. 
\begin{figure}
\includegraphics{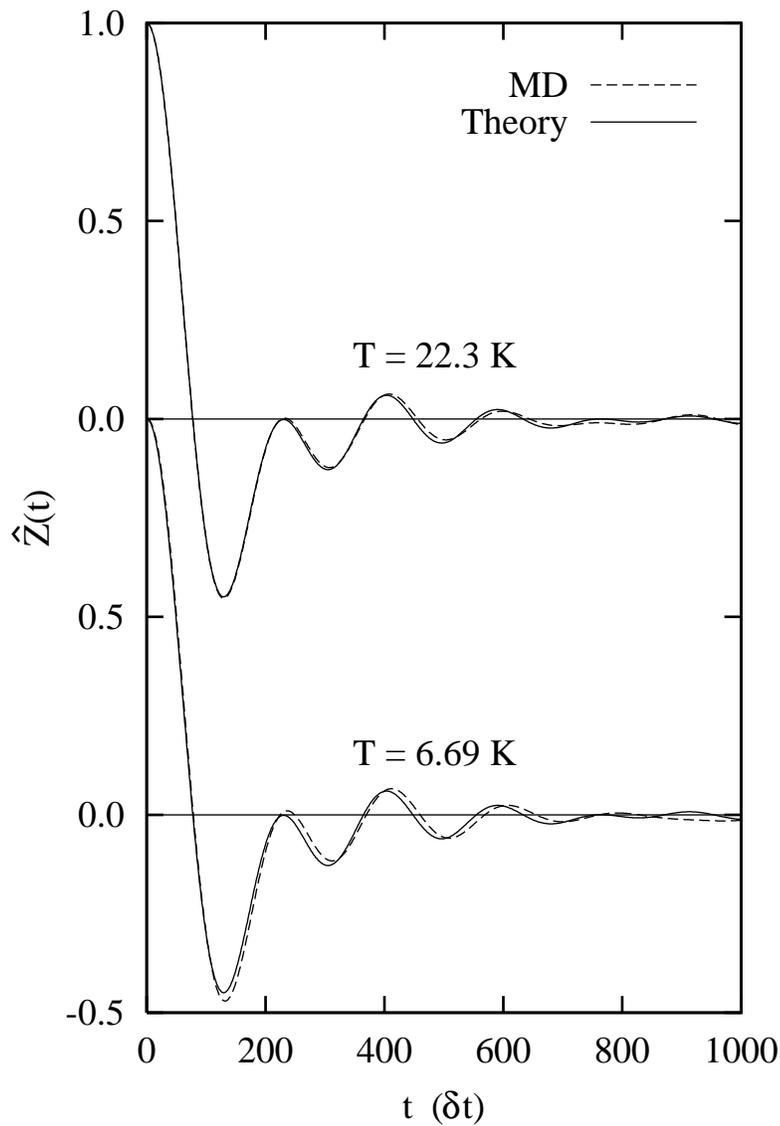}
\caption{The theoretical prediction (Eq.\ (\ref{Zhat})) for $\hat{Z}(t)$ of 
         sodium moving in a single random valley compared with MD data at $T = 
         6.69$ K and $T = 22.3$ K.}
\label{Zvst}
\end{figure}

\section{Conclusions}
\label{concl}

These results show that the motion of a liquid in a single potential valley is 
harmonic to an extremely high approximation; the harmonic prediction for the 
function $\hat{Z}(t)$ matches the calculation from MD very closely.  Any 
contributions due to anharmonicity (which are certainly present) are at most of
the same order as the accuracy of the MD calculations.

Some form of harmonic approximation, such as the one used here, has been taken 
up by many workers attempting to understand the dynamics of liquids, and it is 
helpful to compare their models with our approach.  One of the most popular is 
the theory of Instantaneous Normal Modes (INM), introduced by Rahman, Mandell, 
and McTague \cite{RMM} and LaViolette and Stillinger \cite{LVS} and developed 
extensively by Stratt (for example, \cite{Stratt}).  Stratt expands the 
many-body potential in the neighborhood of an arbitrary point to second order 
in displacements from that point, and he expresses the potential as a quadratic
sum of normal modes, in which the frequencies may be either real or imaginary. 
He then replaces the frequencies by their thermal averages over the potential 
surface, resulting in a temperature-dependent density of frequencies.  From 
this point he calculates the system's motion and considers various time 
correlation functions, including $Z(t)$.  He observes that his \mbox{results} 
are accurate to order $t^{4}$ for short times, but his predictions also diverge
from MD results very rapidly, in a time shorter than half of one vibrational 
period.  The agreement with MD at long times can be improved by omitting the 
imaginary frequencies from the calculation of $Z(t)$, but of course this makes 
the short time behavior inexact.  (The work of Vallauri and Bermejo \cite{VB} 
follows Stratt's procedure.)  Efforts to improve the long time behavior of the 
correlation functions calculated using INM have been made by Madan, Keyes, and 
Seeley \cite{MKS}, who have attempted to extract from the imaginary part of the
INM spectrum a damping factor for $Z(t)$ of the general type suggested by 
Zwanzig \cite{Zwan}.  Also taking their cue from Zwanzig, Cao and Voth 
\cite{CV} have followed a slightly different path, replacing the actual 
potential by a set of temperature-dependent effective normal modes which, as 
they emphasize, bears little resemblance to the mechanical normal modes of a 
single many-particle valley.  In fact, they state quite explicitly that a 
theory based on purely mechanical normal modes will have little success in 
accounting for equilibrium or dynamical properties of liquids.

An obvious difference between our theory and INM is the nature of our 
approximation.  In INM, one approximates the potential quadratically at an 
arbitrary point, with the result that the motion so predicted is accurate only 
for very small times; in our theory, we expand the potential only at very 
special points where we know the predicted motion will be valid for very long 
times.  Both theories then face the problem of extending their validity beyond 
the initial approximation, of course, and we will briefly mention our extension
in the final paragraph below, but there is one particular reason why we 
strongly prefer the approach taken here:  The other models all replace the true
potential by a temperature-dependent potential determined by one or another 
thermal averaging process.  A temperature-dependent potential does not provide 
a true Hamiltonian, and therefore it cannot be used to calculate the quantum or
classical motion, i.e., it cannot be used in the Schr\"{o}dinger equation or 
Newton's law.  (On the dynamical level, temperature is not even a well-defined 
concept.)  Further, the Hamiltonian resulting from a temperature-dependent 
potential cannot be used to do statistical mechanics, except through 
uncontrolled self-consistent procedures.  We prefer to build our theory in 
terms of the actual potential, hence in terms of its true Hamiltonian, and to 
find at least approximate solutions for the Hamiltonian motion, so we can apply
the standard procedures of equilibrium and nonequilibrium statistical 
mechanics.

Further, we would argue that Cao and Voth's skepticism regarding purely 
dynamical approaches is unfounded, given the results here.  It is difficult to 
compare our $\hat{Z}(t)$ results to those of others, because their MD-simulated
states are not always characterized as diffusing or nondiffusing.  We are 
fairly confident that Vallauri and Bermejo's Fig.\ 2b is a comparable state 
(glassy Cs at 20 K), and we believe our fit to MD is slightly better.  Madan, 
Keyes, and Seeley's Fig.\ 3b is an ambiguous case (it is likely that a glass 
transition has occurred), but there also we are confident that our match with 
MD is better.  Hence we would claim that this method shows as much promise as 
the others currently available, and with the physical potential as opposed to a
thermal average potential.

It is also instructive to compare the results of this paper with a model for 
$\hat{Z}(t)$ previously proposed by Wallace \cite{oldvacf} in which a single
particle oscillates in a three-dimensional harmonic valley, and at each turning
point it may with probability $\mu$ ``transit'' to an adjacent valley.  To 
apply that model to a nondiffusing case, we set $\mu = 0$ (indicating no 
transits), yielding $\hat{Z}(t) = \cos(\omega t)$.  Clearly this would not fit 
the MD data for any $\omega$, and it is easy to see why:  Wallace included only
one frequency in his earlier model, whereas our Eq.\ (\ref{Zhat}) contains 
contributions from many frequencies, all of which are necessary to raise the 
first minimum in $\hat{Z}(t)$ above $-1$ and then damp $\hat{Z}(t)$ out by 
dephasing.  This suggests an alternate path to understanding diffusing states: 
Begin with a mean atom trajectory model that by construction reproduces the 
correct result for $\hat{Z}(t)$ in the nondiffusing regime (Eq.\ (\ref{Zhat})),
and then incorporate Wallace's notion of transits into this model.  Our work in
this direction, with comparison to MD data for higher-temperature diffusing 
states of liquid sodium, will be described in a subsequent paper \cite{chis}.


\end{document}